\documentclass[12pt]{article}
\usepackage{amsxtra,amssymb,amsthm,amsmath,latexsym}

\theoremstyle{plain}
\newtheorem{definition}{Definition}[section]

\newtheorem{lemma}{Lemma}[section]

\newtheorem{remark}{Remark}[section]

\def\R{{\mathbb R}}

\def\oH{\buildrel\circ\over H}
\def\oH1{\buildrel\circ\over H\kern-.02in{}^1}

\def\l{\ell}

\def\const{\hbox{\,const\,}}

\begin{document}


\title{Numerically efficient version of the T-matrix method.
}

\author{
A.G. Ramm\\
 Mathematics Department, Kansas State University, \\
 Manhattan, KS 66506-2602, USA\\
ramm@math.ksu.edu\\
}

\date{}

\maketitle\thispagestyle{empty}

\begin{abstract}
A version of the projection method for solving the scattering problem
for acoustic and electromagnetic waves is proposed and shown to be more
efficient numerically than the earlier ones.
\end{abstract}


\section{Introduction}
Consider the scattering problem:
$$ \left( \nabla^2 + k^2 \right) u =0 \hbox{\ in\ } D^\prime :=
    \R^2 \backslash D,
   \eqno{(1.1)}$$
$$u =0 \hbox{\ on\ } S := \partial D,
  \eqno{(1.2)} $$
$$u= e^{ik \alpha \cdot x} + v,
  \eqno{(1.3)}$$
where $D$ is a bounded domain in $\R^2 $ with a piecewise
smooth boundary $S$, and the scattered field
$v$ has the following asymptotics:
$$ v= A(\alpha^\prime, \alpha, k) \frac{e^{ir}}{\sqrt{r}} + o
  \left(\frac{1}{\sqrt{r}} \right), \quad
  r = |x| \to \infty, \frac{x}{r} = \alpha^\prime \in S^1,
  \eqno{(1.4)}$$
and $S^1$ is the unit sphere in $\R^2$. The coefficient $A$ is called the
scattering amplitude, 
$k = \const > 0$ is fixed, and $u:=u(x, \alpha, k)$ is
called the scattering solution.

Let $g(x,y) := \frac{i}{4} H_0^{(1)} (k |x-y|)$.
Take $k=1$ in what follows without loss of generality. Then
$$\left(\nabla^2 + 1 \right) g = - \delta (x-y) \hbox{\ in\ } \R^2,
  g = c \frac{e^{i|y|}}{\sqrt{|y|}} e^{-iy^0 \cdot x} + o
  \left(\frac{1}{\sqrt {|y|}} \right) \hbox{\ as\ } |y| \to \infty,
  y^0 := \frac{y}{|y|}, \eqno{(1.5)}$$
$c = \const = \frac{i}{2\sqrt{2 \pi }}$.
One has, using Green's formula:
$$u(x, \alpha) = e^{i \alpha \cdot x} - \int_S g(x,s) u_N (s^\prime,
\alpha)
  ds^\prime, \quad x \in D^\prime,\, \, k=1,
  \eqno{(1.6)}$$
where $u_N$ is the normal derivative of $u$ on $S$, $N$ is the exterior
unit normal to $S$.

Taking $x \in D^\prime$ to $S$ and using (1.2), one gets
$$Th := f, \quad h:= u_N, \quad f := e^{i \alpha \cdot s},\,\, k=1,
  \eqno{(1.7)}$$
$$Th := \int_S g(s,s^\prime) h(s^\prime) ds^\prime.
  \eqno{(1.8)}$$
Equation (1.7) is the basic equation studied in this paper.

The T-matrix method is described in \cite{R}, \cite{V} and analyzed in
\cite{R} mathematically.

The purpose of this paper is to give a version of this method for solving
equation (1.7), and to analyze this version from the computational
points of view. This is done in sections 2 and 3.

Let $H^\l = H^\l (S)$ be the Sobolev spaces, $\l = 0, 1,$ and
$$Qh : = \int_S \frac{1}{2 \pi} \ln \frac{a}{r_{ss^\prime}} h(s^\prime)
  ds^\prime, \eqno{(1.9)}$$
where $a = \const = \hbox{\ diam\ } D > 0$, so that
$\inf_{s,s^\prime \in S} \frac{a}{r_{ss^\prime}} \geq 1.$

Write (1.7) as
$$Qh + Kh = f, \ f \in H^1, \quad Kh := - \frac{\ln a}{2 \pi} \int_S
  h(s^\prime) ds^\prime + \int_S \left[ g(s,s^\prime) - \frac{1}{2 \pi}
  \ln \frac{1}{r_{ss^\prime}} \right] h dy \eqno{(1.10)}$$

The operator $A = Q^{-1}K$ is compact in $H^1$, and the operator $Q$ is
an isomorphism between $H^0$ and $H^1$ (see Lemmas 1.1 and 1.2 below).
Moreover, $Q$ is a selfadjoint compact positive operator, in
$H^0 : (Qu,u) > 0$ if $u \neq 0$. In Remark 2.1 below we show a possible
usage of (1.10). Let $\{\varphi_j\}_{j=1,2, \dots}$ be a Riesz basis of
$H^0$, that is, every element $u \in H^0$ is uniquely representable
as a convergent in $H^0$ series
$$u = \sum^\infty_{j=1} c_j \varphi_j,
  \eqno{(1.11)}$$
and
$$m \sum^\infty_{j=1} |c_j|^2 \leq \| u \|^2_0 \leq M \sum^\infty_{j=1}
  |c_j|^2, 0 < m \leq M,\quad  m,M = \const>0.
  \eqno{(1.12)}$$

Let us prove the following Lemma 1.1:

\begin{lemma} 
If $k^2$ is not a Dirichlet eigenvalue of the Laplacian in $D$ then the
operator $T$ defined by (1.8) is an isomorphism of $H^0$ onto
$H^1$.
\end{lemma}

\begin{proof}
This result is established in \cite{R} so we only indicate the basic points
of the proof. The assumption of Lemma 2.1 implies the injectivity of
$T:$ if $Th =0$ then the function
$w(x) := \int_S g(x, s^\prime) h (s^\prime) ds^\prime$
solves the Dirichlet problem for the Helmholtz operator
 in $D$ and in $D^\prime,$ and satisfies the
radiation condition at infinity. Thus $w = 0$ in $D$ by the assumption of
Lemma 2.1, and $w = 0$ in $D^\prime$ by lemma 1 in \cite{R}, p.25.
Therefore $h = u_N^+ - u_N^- = u$, where we have used the
jump relation for the normal derivative of the single-layer potentials.
The operator $T : H^0 \to H^1$ is of Fredholm-type: it can be written as
$T = Q + ,K$ where $Q : H^0 \to H^1$ is an isomorphism and
$K$ is compact as an operator from $H^0$ into $H^1$. The injectivity of
$T$ together with its Fredholm property imply the conclusion of Lemma 1.1.
\end{proof}

We assume throughout this paper that $k^2=1$ 
 is not a Dirichlet eigenvalue of
the Laplacian in $D$.

In the literature \cite{V} one usually means by the T-matrix approach (in
acoustic and electromagnetic wave scattering theory) a projection method
for solving equations of the type (1.7) with the following choices of the
basis functions: $\varphi_m = e^{im \theta} H_m (k r (\theta))$ or
$\varphi_m = e^{im \theta} J_m (kr(\theta))$.

These choices lead to the following difficulties discussed in \cite{R}:
as the number $J$ of these functions grows:  $J \to \infty,$ the condition
number of matrix $a_{ij}$ in (2.2)
(see below) grows exponentially and depends strongly on the geometry
of $S$. {\it In contrast,
in our version of the method, the condition number of $a_{ij}$ remains
bounded as $J \to \infty$.}

\begin{lemma} 
The operator $Q^{-1} K$ is compact in $H^1$.
\end{lemma}

\begin{proof}
The kernel of the operator $K$, defined by (1.10), and its first derivatives
are continuous functions of $s$ and $s^\prime$ running through bounded sets,
including the diagonal $s=s^\prime$. By Lemma 1.1 the action of $Q^{-1}$
is equivalent (up to the terms preserving smoothness) to taking the first
order derivatives. Therefore the conclusion of Lemma 2.1 follows.
\end{proof}

\begin{remark} 
Let us outline a possible way to use the splitting in equation (1.10). The
idea is simple: $Q \geq c > 0$ is positive definite. Write  (1.10) as
$$ h+ Ah = F, \quad F : Q^{-1} f. \eqno{(1.13)}$$
Inverting $Q$ numerically is a relatively easy problem since $Q$ is
positive definite. Using an orthonormal basis $\{\varphi_j\}$ of $H^0$,
one can write the projection method for (1.13), namely: 
$$h_J = \sum^J_{j=1} c_j \varphi_j,$$
$$h_i + \sum^J_{j=1} A_{ij} h_j = F_i, \quad
  1 \leq i \leq J, \eqno{(1.14)}$$
where $A_{ij} := (A\varphi_j, \varphi_i)$.
\end{remark}

The matrix in (1.14) is
$B_{ij} = \delta_{ij} + A_{ij}$. If
$\|A\|_0 < 1$, then system (1.4) can be solved by iterations numerically
efficiently. This happens if $k a << 1$, where $a = \hbox{\ diam\ } D$,
but it may happen when the above condition does not hold.

\section{Solution of the basic equation} 
Let us look for an approximate solution to (1.7)
$$h = \sum^J_{j=1} c_j \varphi_j, \eqno{(2.1)}$$
where $\varphi_j$ is a Riesz basis of $H^0$, and
$$\sum^J_{j=1} a_{ij} c_j = f_i \quad 1 \leq i \leq J, \eqno{(2.2)}$$
where
$$f_i := (f, T\varphi_i), \quad a_{ij} := (T \varphi_j, T \varphi_i)_0,
  \eqno{(2.3)}$$
and $T$ is defined in (1.8). Since $T$ is injective, the elements
$\{ T\varphi_j\}$ are linearly independent. Therefore
$$\det (a_{ij})_{1 \leq i,j \leq J} \neq 0 \quad \forall J =1,2,\dots
  \eqno{(2.4)}$$

\begin{definition} 
A system $\{\psi_j\}$ is a Riesz basis of a Hilbert space $H$ if there is
an isomorphism $B$ of $H$ onto $H$ such that $B\psi_j = e_j$, where
$\{e_j\}$ is an orthonormal basis of $H$.
\end{definition}

One gets system (2.2) by solving the following minimization problem:
$$\| \sum^J_{j=1} c_j T \varphi_j - f \|_1 = \min. \eqno{(2.5)}$$
Denote by $\{c_j^{(J)}\}_{1 \leq j \leq J}$ the unique solution to (2.2)
or, equivalently,  to (2.5).

\begin{lemma} 
$\{T \varphi_j\}_{1 \leq j < \infty}$ is a Riesz basis of $H^1$ if
the system
$\{\varphi_j\}_{1 \leq j < \infty}$ is a Riesz basis of $H^0$.
\end{lemma}

\begin{proof}
Let $f \in H^1$ be an arbitrary element of $H^1$. Denote
$T^{-1} f:= h \in H^0$.

Since $\{\varphi_j\}$ is a basis of $H^0$, one has
$$h= \sum^\infty_{j=1} c_j \varphi_j, \quad
  f = \sum c_j T \varphi_j.$$
If $ \sum c_j T \varphi_j=0, $ then, applying the continuous operator
$T^{-1}$, one gets $\sum^\infty_{j=1} c_j \varphi_j = 0$, so $c_j = 0$
for all $j$ since $\{\varphi_j\}_{1 \leq j < \infty}$ is a basis of
$H^0$. We have proved that $\{T\varphi_j\}_{1 \leq j < \infty}$
is a basis of $H^1$.

Let us prove that if $\{\varphi_j\}_{1 \leq j < \infty}$ is a Riesz basis
of $H^0$ then $\{T \varphi_j\}_{1 \leq j < \infty}$
is a Riesz basis of $H^1$, that is, there exists an isomorphism $B$ of
$H^1$ onto $H^1$ such that $BT \varphi_j = e_j$,
$(e_j, e_i)_1 = \delta_{ij} =\begin{cases} 1 &i =j \\ 0 &i \neq j \end{cases}.$

Let $\{e_j\}$ be an orthonormal basis of $H^1$. Define a linear operator
$F: H^1 \to H^1$ by the formula:
$$F \sum^\infty_{j=1} c_j e_j := \sum^\infty_{j=1} c_j
  T \varphi_j,$$
in particular, $F e_j = T \varphi_j$. Let us prove that $F$ is an
isomorphism of $H^1$ onto $H^1$. If this is proved, then $B:= F^{-1}$, and
Lemma 2.1 is proved.

Clearly $F$ is linear, is defined on all of $H^1$, and is continuous. Only
the continuity of $F$ needs a proof.

Let $u_n \to u$ in $H^1$. Then
$u_n = \sum^\infty_{j=1} c_j^{(n)} e_j, \quad u = \sum^\infty_{j=1} c_j e_j,
 \quad \sum^\infty_{j=1} |c_j^{(n)} - c_j|^2 \to 0$,
 $Fu_n = \sum^\infty_{j=1} c_j^{(n)} T\varphi_j, \quad  Fu =
   \sum^\infty_{j=1} c_j T \varphi_j$.
Thus:
$$
\begin{aligned}
  \| Fu_n - Fu \|_1^2 = \| \sum^\infty_{j=1} (c_j^{(n)} - c_j) T \varphi_j
  \|^2_1 = \\
  \|T \sum^\infty_{j=1} (c_j^{(n)} - c_j) \varphi_j \|^2_1 \leq
  \| T \|^2 \| \sum^\infty_{j=1} (c_j^{(n)} - c_j) \varphi_j \|_0 \leq \\
  \| T \|^2 M \sum^\infty_{j=1} |c_j^{(n)} - c_j|^2 \to 0\quad \hbox{\ as\
}
  n \to \infty,
  \end{aligned}$$
where we have used the assumption that $\{\varphi_j\}_{1 \leq j < \infty}$
a Riesz basis of $H^0$.

Thus $F$ is a linear continuous, defined on all of $H^1$ operator.

Therefore $F$ is bounded.

Let us check that $F$ is injective: if $u \in H^1$,
$u = \sum^\infty_{j=1} c_j e_j$, and $Fu=0$ then
$\sum^\infty_{j=1} c_j T \varphi_j =0$. Apply $T^{-1}$ and get
$\sum^\infty_{j=1} c_j \varphi_j = 0$. Thus, $c_j = 0$ $\forall j$, since
$\{\varphi_j\}$ is a basis. The injectivity of $F$ is proved.

To complete the
proof one has to check that the range of $F$ is the whole space $H^1$. Let
us do this. Take an arbitrary $f \in H^1$ and define $h := T^{-1} f \in
H^0$.
Let $h = \sum^\infty_{j=1} c_j \varphi_j$, then
$$Th = f = \sum^\infty_{j-=1} c_j T \varphi_j = F \sum^\infty_{j=1}
  c_j e_j.$$
Therefore $F$ is an isomorphism of $H^1$ onto $H^1$, and
$\{T \varphi_j\}$ is a Riesz basis of $H^1$, as claimed. Lemma 2.1 is proved.
\end{proof}

Let us summarize the proposed method for solving the basic equation (1.7):

{\it Step 1.} {\it Choose a Riesz basis $\{\varphi_j\}_{1 \leq j < \infty}$
in $H^0 = L^2(S)$.}

We discuss this choice below.

{\it Step 2.} {\it Calculate the matrix entries $a_{ij}$ and the numbers
$f_i, \quad 1 \leq i, j \leq J$, where $J$ is an a priori chosen integer.}

{\it Step 3.}
{\it Solve linear system (2.2) numerically.}

{\it The matrix in (2.2) has condition number that remains bounded when
$J$
grows, as follows from Lemma 2.1.}

Let us discuss the choice of the basis $\{\varphi_j\}$.

Assume that $r = r(\theta)$ is the equation of $S$ in the two-dimensional
case that is, $S$ is star-shaped. The element of the arc length of $S$ is
$ds = \sqrt{r^{'2} (\theta) + r^2(\theta)} d \theta := a
(\theta) d\theta$.
Let $S^1$ denote the unit sphere
$S^1 := \{x : x \in \R^2, |x| = 1\}$. Choose
$$\varphi_{co} (s) = \frac{1}{\sqrt{2 \pi a (\theta)}},\quad \varphi_{cm}
(s) :=
  \frac{\cos (m \theta)}{\sqrt{\pi a (\theta)}} \quad
  \varphi_{sm} (s) = \frac{\sin (m \theta)}{\sqrt{\pi a (\theta)}}, \quad 
  m = 1,2, \dots \eqno{(2.6)}$$
where $s = (r (\theta), \theta)$.

Then
$$\int_S \varphi_{cm} (s) \varphi_{sm^\prime} (s) ds = \frac{1}{\pi}
  \int^{2 \pi}_0 \cos (m \theta) \sin (m^\prime \theta) d \theta = 0,$$
and
$$\int_S \varphi_{cm} \varphi_{cm^\prime} ds = \delta_{mm^\prime}, \quad
  \int_S \varphi_{sm} \varphi_{sm^\prime} ds = \delta_{mm^\prime},$$
so that $\{\varphi_m (s)\}$ is not only a Riesz basis of
$H^0 = L^2(S)$, but an orthonormal basis of $H^0$. 

Similar
construction holds in $\R^3$, where the normalized spherical harmomics
$Y_{\l m} (\alpha)$ are used in place of
$\cos (m \theta)$ and $\sin (m \theta)$, $\alpha = (\theta, \varphi)$
is the unit vector in $\R^3, \alpha \in S^2, S^2$ is the unit sphere in
$\R^3$.


\begin{thebibliography}{10}

\bibitem{R}
Ramm, A.G.,
Scattering by Obstacles, D. Reidel, Dordrecht, 1986.

\bibitem{V}
Varadan, V.K., Varadan, V.V., (editors),
Acoustic, Electromagnetic and Elastic Wave Scattering- Focus on the
T-matrix approach, Pergamon Press, New York, 1980.





\end{thebibliography}
\end{document}